\documentclass[runningheads]{llncs}
\usepackage{graphicx}
\pdfoutput=1

\usepackage{rotating}
\usepackage{multirow}
\usepackage{longtable}
\usepackage{bm}
\usepackage{bbm}

\usepackage{hyperref}
\usepackage[table,xcdraw]{xcolor}
\usepackage[vlined,boxed]{algorithm2e}
\SetAlgoCaptionSeparator{.}

\SetCommentSty{mycommfont}

\SetAlCapSkip{1em}
\usepackage[section]{placeins}
\usepackage{amsmath}
\usepackage{amssymb}
\usepackage{float, caption}
\usepackage{pdfpages}
\usepackage[colorinlistoftodos]{todonotes}
\usepackage{siunitx}

\begin{document}
\title{Automated Retrieval of ATT\&CK Tactics and Techniques for Cyber Threat Reports}
\titlerunning{Automated Retrieval of TTPs from Threat Reports}

\author{
Valentine Legoy\inst{1} \and
Marco Caselli\inst{2} \and
Christin Seifert\inst{1} \and 
Andreas Peter\inst{1}}

\authorrunning{V. Legoy et al.}

\institute{University of Twente, 7522 NB Enschede, The Netherlands\\
\email{\{c.seifert, a.peter\}@utwente.nl}\\ \and
Siemens Corporate Technology, 81739 Munich, Germany\\
\email{marco.caselli@siemens.com}}

\maketitle            
\begin{abstract}
Over the last years, threat intelligence sharing has steadily grown, leading cybersecurity professionals to access increasingly larger amounts of heterogeneous data. Among those, cyber attacks' Tactics, Techniques and Procedures (TTPs) have proven to be particularly valuable to characterize threat actors' behaviors and, thus, improve defensive countermeasures. Unfortunately, this information is often hidden within human-readable textual reports and must be extracted manually. In this paper, we evaluate several classification approaches to automatically retrieve TTPs from unstructured text. To implement these approaches, we take advantage of the MITRE ATT\&CK framework, an open knowledge base of adversarial tactics and techniques, to train classifiers and label results.
Finally, we present rcATT, a tool built on top of our findings and freely distributed to the security community to support cyber threat report automated analysis.
\end{abstract}
\section{Introduction}
According to Gartner, Cyber Threat Intelligence (CTI) can be generally identified as any piece of information corresponding to ``evidence-based knowledge (...) about an existing or emerging menace''\cite{gartnerthreatintelligence}. Despite lacking a precise definition that comprehensively describes its properties, the security community generally agrees on differentiating among three main types of CTI: ``tactical'', ``operational'' and ``strategical''.
Tactical CTI refers to information that is a direct manifestation of adversary actions within a compromised system; examples are IP addresses, file hashes and other artifacts that are traces of actual malicious activities. Operational CTI relates to more general information that has an impact on day-to-day security decision making; for instance statistics of ongoing cyberattack campaigns requiring security teams to deploy specific countermeasures. Finally, strategic CTI refers to knowledge about threat actors' capabilities and motivations, such as characteristics and behavior of advanced persistent threats.

Over the last year, the sharing of these three types of CTI has steadily grown. Threat information sources have increased in number, supported by the development of standardized technologies and formats, like STIX~\cite{barnum2012standardizing}. Available information is either open source and freely accessible over the Internet or selectively provided within private feeds managed by specialized companies, like IBM X Force~\cite{ibmxforce2019} and OTX AlienVault~\cite{alienvaultotx2019}.

It is worth noting that the amount of shared information is not equally distributed among the three types of CTI. In fact, most sharing activities involve tactical CTI in the form of Indicators of Compromise (IOCs). Malicious IPs and URLs as well as signatures of malware executables cover the vast majority of CTI exchanges while operational and, especially, strategic CTI corresponds to almost negligible percentages despite their importance~\cite{dog2016strategic}. Furthermore, operational and strategic CTI does not usually come within structured formats but is embedded within human-readable cyber threat reports (CTRs).
If structured data can be generally handled simply by using web scrapers and parsers, unstructured data almost always requires security experts to read and manually extract the most relevant information. Considering the number of CTRs published each day,\footnote{IBM X Force and OTX AlienVault alone have an average of 15 new reports per day.} dealing with unstructured security information currently remains a cumbersome task.

Our work aims to overcome this challenging task by simplifying the extraction of valuable security information contained in CTRs, namely cyber threats' Tactics, Techniques and Procedures (TTPs). To achieve this, we use the ATT\&CK (Adversarial Tactics, Techniques, and Common Knowledge) framework~\cite{mitreattack}, an open knowledge base of adversarial tactics and techniques, and we aim at labelling each CTR with the TTPs that most likely match its content. Automating this labeling allows security experts to use the information within any given CTR more efficiently and thus support the definition of prevention, detection and mitigation methods for the related threats. Our contribution specifically focuses on three main aspects:
\begin{itemize}
    \item We implement and evaluate standard multi-label text classification models and adapt them to retrieve ATT\&CK tactics and techniques from CTRs;
    \item We test several post-processing techniques based on the hierarchical structure of the ATT\&CK framework to enhance classification results;
    \item We present and distribute {\it rcATT},\footnote{\url{https://github.com/vlegoy/rcATT}} a tool based on our findings that: 1.~predicts ATT\&CK tactics and techniques related to a given CTR, 2.~allows the user to correct predictions and give feedback on the classification engine to improve results over time, and 3.~outputs the results using the Structured Threat Information eXpression (STIX) format.
\end{itemize}
Finally, we provide a comparison with available open-source solutions showing rcATT's improvements over the state of the art.
\section{ATT\&CK framework}\label{framework}

MITRE ATT\&CK (Adversarial Tactics, Techniques, and Common Knowledge) framework~\cite{mitreattack} is a ``globally-accessible knowledge base of adversary tactics and techniques based on real-world observations''. Its comprehensive collection of tactics and techniques has gained popularity over recent years and has been integrated into popular threat information sharing technologies~\cite{barnum2012standardizing}.

ATT\&CK provides a structured taxonomy to describe several different adversary behaviours. It formally divides into three ``technology domains":
\begin{itemize}
\item ``Enterprise", which describes behaviours on standard IT systems (e.g., Linux, Windows).
\item ``Mobile", which focuses on mobile devices  (e.g., Android, iOS). 
\item ``ICS'', which relates to industrial control and, more in general, to cyber-physical systems
\end{itemize}
Beyond these domains, ATT\&CK also documents behaviours for reconnaissance and weaponisation under the ``PRE-ATT\&CK" designation. Although our work applies likewise to every domain, we will focus this paper on the ATT\&CK ``Enterprise''.

The ``Enterprise'' ATT\&CK framework is usually represented as a matrix of tactics and techniques where:
\begin{itemize}
\item the tactics represent possible goals of an attacker (e.g., ``Initial Access", ``Privilege Escalation", etc.)
\item the techniques identify ``how" an attacker might fulfill a specific goal (e.g., ``Access Token Manipulation'', ``Accessibility Features'', etc.)
\end{itemize}

Tactics and techniques are the key focus of our work as they will correspond to the labels of our classification.

Each technique is associated with one or more tactics. Furthermore, the ``Enterprise'' ATT\&CK framework collects information about recorded adversary use of every available technique (e.g., involved threat actors, notorious malware, etc.) and possible mitigation approaches. All this information will be used to train our classifiers.

\subsection{Data Characterization}\label{collection}
The actual ATT\&CK data employed in our work is stored on GitHub\footnote{\url{https://github.com/mitre/cti/tree/master/enterprise-attack}, accessed February 2020} and represented using the STIX 2.0 format~\cite{barnum2012standardizing}.

\begin{figure}[!htb]
        \center{\includegraphics[width=\columnwidth]
        {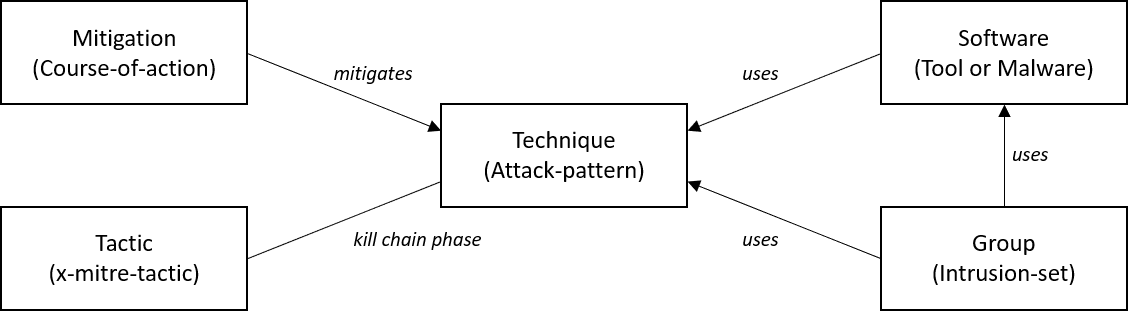}}
        \caption{\label{fig1} ``Enterprise'' ATT\&CK data schema and corresponding STIX 2.0 objects}
\end{figure}

Figure~\ref{fig1} shows the structure of the employed dataset and emphasizes the relationships 
linking all available information to the related tactics and techniques.
Every piece of information includes several external references (e.g., usually threat reports or technical descriptions). All text linked by these references was considered part of the dataset and used to train the classifiers. 

Overall, the ``Enterprise'' ATT\&CK dataset is enriched by 1490 different security reports either in HTML (1311) or in PDF (179) format. We handled and parsed documents in PDF format using state-of-the-art tools~\cite{bast2017benchmark}. To extract content from HTML resources, we focused only on HTML paragraph tags to simplify the process and avoid noise. All available text was eventually checked to remove strings that can hinder the classification (e.g., we removed all stop words provided in the list of the Natural Language Toolkit (NLTK)~\cite{Loper02nltkthe})

In the end, all ATT\&CK tactics linked to at least 80 reports and provide a suitable dataset for the training of the classifiers. Unfortunately, this was not the same for the techniques which presented instead an imbalance (e.g., some techniques corresponding to just one report). We omitted all techniques that had less than 5 reports in the dataset. This ensures that we can at least train on 4 reports and have 1 report for testing in a 5-fold cross-validation setting.

\section{Methodology overview}\label{methodology}
Considering that CTRs may refer to more than one tactic (or technique) at the same time, the machine learning problem is a multi-class multi-label approach. We tested both general approach to multi-label classification: i) dedicated multi-label classifiers and ii) multiple single-label classifiers, each of which is responsible to decide whether its class (e.g., one specific tactic) should be assigned or not.

During the evaluation, we tested several  classifiers simultaneously comparing different text representation methods such as  term-frequency (TF) and term frequency-inverse document frequency (TF-IDF) weighting factors~\cite{manning2008scoring} as well as Word2Vec~\cite{mikolov2013efficient}.
Finally, we defined several post-processing approaches to improve classification results. These approaches are based on properties and characteristics of the ``Enterprise'' ATT\&CK framework. For example, we leverage the hierarchical structure of the framework by filtering the classification results based on a coherent matching of tactics and techniques belonging to those tactics. 
In all experiments, we performed 5-fold cross-validation.

\subsection{Metrics}

Our evaluation metrics of choice are: precision, recall, and $F_{0.5}$ score. The $F_{0.5}$ score represents a weighted average between the precision $\pi$ and recall $\rho$ where the precision is considered twice as important as the recall. The  $F_{0.5}$ is computed as follows:

\begin{align*}
\pi = \frac{TP}{TP+FP},\quad \rho = \frac{TP}{TP+FN},\quad
F_{0.5} = 1+ 0.5^2\cdot \frac{\pi \cdot \rho}{0.5^2\pi+\rho}
\end{align*}
with TP being true positives, FP false positives and FN false negatives.
We chose $F_{0.5}$ to emphasise precision over recall, i.e., the assigned labels need to be correct (precision), but there might be labels we do not assign (recall). Finally, we employ macro-averaging and micro-averaging~\cite{tsoumakas2009mining}. 

\subsection{Baseline}
To simplify the analysis of different classification models and quickly identify unsuccessful ones, we defined a ``naive'' baseline by simply attributing to each testing instance the most frequent label of the training set. Table~\ref{tab4} shows the results of this baseline.

\begin{table}[!htb]
\caption{Naive baseline based on the attribution of the most frequent label}
\centering
\resizebox{9.5cm}{!}{%
\begin{tabular}{|l|c|c|c|c|c|c|}
\hline
\multirow{2}{*}{Majority} & \multicolumn{3}{c|}{Micro}                         & \multicolumn{3}{c|}{Macro}                     \\ \cline{2-7} 
                                & \multicolumn{1}{l|}{Precision} & Recall  & $F_{0.5}$    & Precision & Recall & \multicolumn{1}{l|}{$F_{0.5}$} \\ \hline
Tactics                         & 48.72\%                        & 19.00\% & 37.10\% & 4.43\%    & 9.09\% & 4.93\%                    \\ \hline
Techniques                      & 9.57\%                         & 2.51\%  & 6.11\%  & 0.05\%    & 0.48\% & 0.06\%                    \\ \hline
\end{tabular}%
}
\label{tab4}
\end{table}

\section{Multi-label classification}\label{initialclass}

As introduced in the previous section, our evaluation consists on testing different text representation methods with several multi-label classification models. 

\subsection{Text representation}
Our primary approach is based on standard weighting functions such as term-frequency (TF) and term frequency-inverse document frequency (TF-IDF)~\cite{manning2008scoring}. These weighting functions were tested employing either simple bag-of-words or bi-grams and tri-grams. The use of TF and TF-IDF weighting functions relied on the application of a maximum and a minimum frequency of appearance in the overall corpus.
According to the performed tests, a simple bags-of-words representation performed better than any other grouping technique. Furthermore, decreasing the number of features and thus selecting half of the words with highest TF/TF-IDF scores outperformed both the choice of a constant number of words for each report or not limiting the number of features at all.

Besides TF and TF-IDF, our analysis includes tests based on Word2Vec~\cite{mikolov2013efficient}. In this case, Word2Vec was trained on our dataset instead of taking advantage of the existing pre-trained versions. This was due to some inconsistencies found in some early results. Tests with Word2Vec were performed both averaging and summing word vectors representing the text.

\subsection{Classification models}

Our analysis includes two main strategies to tackle the multi-label classification problem: adapting state-of-the-art classification algorithms or approaching the overall problem from a different perspective. In this last case, the approaches are: using binary relevance and training a binary classifier for each label independently~\cite{luaces2012binary}, employing a classifier chain (which is similar to binary relevance but uses the relation between labels~\cite{read2011classifier}), or adopting label  power sets (basically transforming them into a multi-class problems between all labels combination~\cite{spolaor2013comparison}). A power-set model is, however, difficult to apply to our case, as we have too many labels and not enough data to cover all possible combinations. Therefore, we only focus on binary relevance and classifier chains and test different types of classifiers.

All implementations of multi-label classifiers are built on top of the Scikit-learn library (version 0.21.3)~\cite{sklearnapi2019}. Our tests include: multi-label K-Nearest Neighbours, multi-label Decision Tree and Extra Tree techniques, and Extra Trees, Random Forest ensemble methods for multi-label classification. For what concerns binary relevance and classifier chains, we tested several linear models. These are: Logistic Regression, Perceptron, Ridge, Bernoulli Naive Bayes classifier, K-Nearest Neighbors and Linear Support Vector Machine~(Linear SVC). We also included multiple tree-based models such as: Decision Tree, Extra Tree classifiers, Random Forest, Extra Trees, AdaBoost Decision Tree, Bagging Decision Tree and Gradient Boosting. 

It is to be noted that in classification models were the correlations between labels is used (i.e. multi-label classifiers and classifier chains), we predicted the tactics and techniques together, with the best parameters possible for both label types. In the cases in which the label relationship did not matter (i.e. binary relevance), we split the classification by tactics and techniques, applying to each the best parameters possible for their category.

To avoid overfitting, in addition to the previously mentioned reduction of the number of features, we tested regularising models, fine-tuning the hyper-parameters of the classifiers and aim at simple models. 

As an attempt to solve the imbalance in the dataset, we used a random resampling approach on the models with binary relevance, associated with TF and TF-IDF. In the case of the tactics, the resampling method was a combination of over- and undersampling for all tactics to have in their training set with 400 positive reports and 400 negative reports. For the techniques,  we randomly sampled 125 positive reports and 500 negative reports.

\subsection{Results and discussion}

Tables~\ref{tabtactres} and ~\ref{tabtechres} show the performance of the tactics and techniques classification, respectively. Classifiers that are not shown in the table either did not outperform our naive baseline or obtained overall lower results than those present.

\begin{table}[]
\caption{Classification results for tactics prediction. Abbreviations: CC = Classifier Chain, BR = Binary relevance, DT = Decision Tree, T = Tree? KNN = K Nearest Neighbors}
\label{tabtactres}
\centering
\resizebox{\textwidth}{!}{%
\begin{tabular}{|ll|c|c|c|c|c|c|l|c|c|c|c|c|c|}
\cline{1-8} \cline{10-15}
\multicolumn{2}{|l|}{\multirow{3}{*}{}} & \multicolumn{6}{c|}{Without resampling}                                                                                           & \multirow{3}{*}{} & \multicolumn{6}{c|}{With resampling}                                                                                                                \\ \cline{3-8} \cline{10-15} 
\multicolumn{2}{|l|}{}                  & \multicolumn{3}{c|}{Micro}                                      & \multicolumn{3}{c|}{Macro}                                      &                   & \multicolumn{3}{c|}{Micro}                                               & \multicolumn{3}{c|}{Macro}                                               \\ \cline{3-8} \cline{10-15} 
\multicolumn{2}{|l|}{}                  & Precision        & Recall                    & $F_{0.5}$             & Precision        & Recall                    & $F_{0.5}$             &                   & Precision                 & Recall           & $F_{0.5}$                      & Precision                 & Recall           & $F_{0.5}$                      \\ \cline{1-8} \cline{10-15}
\multicolumn{2}{|l|}{Majority Baseline} & 48.72\%          & 19.00\%                   & 37.10\%          & 4.43\%           & 9.09\%                    & 4.93\%           &                   & -                         & -                & -                         & -                         & -                & -                         \\ \hline
\multicolumn{15}{|l|}{Term Frequency}                                                                                                                                                                                                                                                                                                                 \\ \hline
        & CC Adaboost DT                & 64.73\%          & 50.84\%                   & 61.30\%          & 60.87\%          & 45.20\%                   & 56.45\%          &                   & -                         & -                & -                         & -                         & -                & -                         \\ \cline{1-8} \cline{10-15} 
        & CC Bagging DT                 & 67.19\%          & 40.57\%                   & 59.38\%          & 64.01\%          & 33.46\%                   & 51.89\%          &                   & -                         & -                & -                         & -                         & -                & -                         \\ \cline{1-8} \cline{10-15} 
        & CC Gradient T Boosting        & 71.84\%          & 44.33\%                   & 63.61\%          & 66.72\%          & 37.42\%                   & 55.91\%          &                   & -                         & -                & -                         & -                         & -                & -                         \\ \cline{1-8} \cline{10-15} 
        & CC Logistic Regression        & 63.81\%          & 54.35\%                   & 61.61\%          & 58.86\%          & 47.78\%                   & 55.85\%          &                   & -                         & -                & -                         & -                         & -                & -                         \\ \cline{1-8} \cline{10-15} 
        & CC Perceptron                 & 56.34\%          & 56.58\%                   & 56.35\%          & 51.63\%          & 50.79\%                   & 51.04\%          &                   & -                         & -                & -                         & -                         & -                & -                         \\ \cline{1-8} \cline{10-15} 
        & CC Linear SVC                 & 63.81\%          & 51.02\%                   & 60.71\%          & 58.89\%          & 44.31\%                   & 54.64\%          &                   & -                         & -                & -                         & -                         & -                & -                         \\ \cline{1-8} \cline{10-15} 
        & BR AdaBoost DT                & 62.30\%          & 49.27\%                   & 59.08\%          & 57.26\%          & 42.37\%                   & 52.91\%          &                   & 49.77\%                   & 60.45\%          & 47.08\%                   & 52.24\%                   & 65.22\%          & 48.91\%                   \\ \cline{1-8} \cline{10-15} 
        & BR Bagging DT                 & 68.26\%          & 42.36\%                   & 60.75\%          & 63.71\%          & 34.30\%                   & 51.54\%          &                   & 52.39\%                   & 58.59\%          & 49.36\%                   & 54.32\%                   & 63.73\%          & 50.48\%                   \\ \cline{1-8} \cline{10-15} 
        & BR Gradient T Boosting        & 71.42\%          & 48.19\%                   & 65.04\%          & 66.35\%          & 40.16\%                   & 56.78\%          &                   & 54.87\%                   & \textbf{66.41\%} & 51.94\%                   & 57.66\%                   & 72.47\%          & 53.92\%                   \\ \cline{1-8} \cline{10-15} 
        & BR Logistic Regression        & 63.71\%          & 54.42\%                   & 61.54\%          & 58.74\%          & 47.81\%                   & 55.76\%          &                   & 56.94\%                   & 61.56\%          & 52.44\%                   & 58.66\%                   & \textbf{66.74\%} & 53.90\%                   \\ \cline{1-8} \cline{10-15} 
        & BR Ridge Classifier           & 61.85\%          & 51.46\%                   & 59.39\%          & 55.64\%          & 45.31\%                   & 52.92\%          &                   & 49.93\%                   & 55.86\%          & 45.48\%                   & 51.50\%                   & 58.97\%          & 47.03\%                   \\ \cline{1-8} \cline{10-15} 
        & BR Linear SVC                 & 64.70\%          & 51.55\%                   & 61.51\%          & 59.20\%          & 44.36\%                   & 54.89\%          &                   & 54.95\%                   & 57.69\%          & 50.76\%                   & 56.45\%                   & 63.37\%          & 51.83\%                   \\ \cline{1-8} \cline{10-15}
        & Adapted KNN           & 57.93\% & 38.66\% & 52.55\% & 52.55\% & 30.41\% & 43.06\% & & -                         & -                & -                         & -                         & -                & -                    \\ \cline{1-8} \cline{10-15}
\multicolumn{15}{|l|}{Term Frequency-Inverse Document Frequency}                                                                                                                                                                                                                                                                                      \\ \hline
        & CC Adaboost DT                & 61.42\%          & 49.86\%                   & 58.59\%          & 57.71\%          & 44.09\%                   & 53.79\%          &                   & -                         & -                & -                         & -                         & -                & -                         \\ \cline{1-8} \cline{10-15} 
        & CC Gradient T Boosting        & 71.15\%          & 43.15\%                   & 62.52\%          & 67.40\%          & 36.29\%                   & 54.97\%          &                   & -                         & -                & -                         & -                         & -                & -                         \\ \cline{1-8} \cline{10-15} 
        & CC Perceptron                 & 62.06\%          & 54.97\%                   & 60.28\%          & 58.05\%          & 49.26\%                   & 54.72\%          &                   & -                         & -                & -                         & -                         & -                & -                         \\ \cline{1-8} \cline{10-15} 
        & CC Ridge Classifier           & 74.40\%          & 41.27\%                   & 63.27\%          & \textbf{67.63\%} & 33.31\%                   & 51.74\%          &                   & -                         & -                & -                         & -                         & -                & -                         \\ \cline{1-8} \cline{10-15} 
        & CC Linear SVC                 & 71.63\%          & 44.89\%                   & 63.41\%          & 65.70\%          & 36.76\%                   & 54.59\%          &                   & -                         & -                & -                         & -                         & -                & -                         \\ \cline{1-8} \cline{10-15} 
        & BR AdaBoost DT                & 61.02\%          & 51.02\%                   & 58.61\%          & 56.61\%          & 44.67\%                   & 53.19\%          &                   & 49.52\%                   & 58.46\%          & 46.12\%                   & 51.84\%                   & 63.79\%          & 47.86\%                   \\ \cline{1-8} \cline{10-15} 
        & BR Bagging DT                 & 66.88\%          & 41.44\%                   & 59.39\%          & 63.92\%          & 34.95\%                   & 52.29\%          &                   & 53.15\%                   & 56.68\%          & 50.06\%                   & 54.70\%                   & 61.94\%          & 50.87\%                   \\ \cline{1-8} \cline{10-15} 
        & BR Gradient T Boosting        & 70.13\%          & 46.85\%                   & 63.66\%          & 65.08\%          & 38.94\%                   & 55.03\%          &                   & 54.30\%                   & 63.40\%          & 50.79\%                   & 56.92\%                   & 70.53\%          & 52.48\%                   \\ \cline{1-8} \cline{10-15} 
        & BR Logistic Regression        & 71.04\%          & 50.70\%                   & 65.61\%          & 59.00\%          & 40.53\%                   & 51.21\%          &                   & 59.25\%                   & 63.13\%          & \textit{\textbf{56.69\%}} & 61.09\%                   & 69.80\%          & \textit{\textbf{57.14\%}} \\ \cline{1-8} \cline{10-15} 
        & BR Perceptron                 & 65.20\%          & 55.35\%                   & 62.80\%          & 60.54\%          & 48.29\%                   & 56.24\%          &                   & 57.26\%                   & 62.68\%          & 53.97\%                   & 59.28\%                   & 69.11\%          & 54.98\%                   \\ \cline{1-8} \cline{10-15} 
        & BR Ridge Classifier           & \textbf{72.40\%} & 48.90\%                   & \textbf{65.83\%} & 66.57\%          & 38.58\%                   & 53.32\%          &                   & \textit{\textbf{59.47\%}} & 62.28\%          & 55.77\%                   & \textit{\textbf{61.13\%}} & 68.89\%          & 56.59\%                   \\ \cline{1-8} \cline{10-15} 
        & BR Linear SVC                 & 65.64\%          & \textit{\textbf{64.69\%}} & 65.38\%          & 60.26\%          & \textit{\textbf{58.50\%}} & \textbf{59.47\%} &                   & 57.71\%                   & 66.17\%          & 53.88\%                   & 59.97\%                   & 71.18\%          & 55.75\%                   \\ \cline{1-8} \cline{10-15} 
        & Adapted KNN           & 62.45\% & 45.24\% & 57.89\% & 57.82\% & 36.75\% & 49.63\% & & -                         & -                & -                         & -                         & -                & -                    \\ \cline{1-8} \cline{10-15}
        \multicolumn{15}{|l|}{Word2Vec average}                                                                                                                                                                                                                                                                                      \\ \hline
        & CC Adaboost DT                & 58.59\% & 44.21\% & 54.98\% & 52.69\% & 36.95\% & 47.34\%          &                   & -                         & -                & -                         & -                         & -                & -                         \\ \cline{1-8} \cline{10-15} 
        & CC Gradient T Boosting        & 63.33\% & 42.22\% & 57.58\% & 55.10\% & 33.38\% & 47.52\%          &                   & -                         & -                & -                         & -                         & -                & -                         \\ \cline{1-8} \cline{10-15} 
        & CC Logistic Regression                 & 62.80\% & 34.15\% & 53.78\% & 55.27\% & 26.87\% & 42.63\%          &                   & -                         & -                & -                         & -                         & -                & -                         \\ \cline{1-8} \cline{10-15} 
        & CC KNN           & 53.40\% & 51.78\% & 52.93\% & 48.15\% & 43.84\% & 45.79\%          &                   & -                         & -                & -                         & -                         & -                & -                         \\ \cline{1-8} \cline{10-15} 
        & CC Linear SVC                 & 62.88\% & 40.82\% & 56.75\% & 59.98\% & 34.04\% & 49.66\%          &                   & -                         & -                & -                         & -                         & -                & -                         \\ \cline{1-8} \cline{10-15} 
        & BR AdaBoost DT                & 57.97\% & 46.24\% & 55.11\% & 50.77\% & 38.54\% & 46.70\%          &                   & -                         & -                & -                         & -                         & -                & -                    \\ \cline{1-8} \cline{10-15} 
        & BR Gradient T Boosting                 & 64.53\% & 44.77\% & 59.23\% & 55.65\% & 35.07\% & 46.78\%           &                   & -                         & -                & -                         & -                         & -                & -                  \\ \cline{1-8} \cline{10-15} 
        & BR Logistic Regression        & 66.86\% & 41.94\% & 59.68\% & 61.02\% & 31.95\% & 46.17\%          &                   & -                         & -                & -                         & -                         & -                & -                   \\ \cline{1-8} \cline{10-15} 
        & BR KNN        & 58.08\% & 51.48\% & 56.58\% & 51.80\% & 41.86\% & 47.45\% & & -                         & -                & -                         & -                         & -                & - \\ \cline{1-8} \cline{10-15} 
        & BR Linear SVC                 & 64.85\% & 44.91\% & 59.49\% & 56.57\% & 35.35\% & 47.81\% & & -                         & -                & -                         & -                         & -                & -                   \\ \cline{1-8} \cline{10-15} 
        & Adapted KNN           & 57.33\% & 51.03\% & 55.92\% & 52.11\% & 41.78\% & 47.43\%& & -                         & -                & -                         & -                         & -                & -                    \\ \cline{1-8} \cline{10-15}
         \multicolumn{15}{|l|}{Word2Vec sum}                                                                                                                                                                                                                                                                                      \\ \hline
        & CC Adaboost DT                & 60.26\% & 43.49\% & 55.82\% & 54.81\% & 36.87\% & 48.68\%          &                   & -                         & -                & -                         & -                         & -                & -                         \\ \cline{1-8} \cline{10-15} 
        & CC Gradient T Boosting        & 60.21\% & 40.00\% & 54.69\% & 50.19\% & 30.78\% & 43.26\%          &                   & -                         & -                & -                         & -                         & -                & -                         \\ \cline{1-8} \cline{10-15} 
        & CC Logistic Regression                 & 54.83\% & 50.41\% & 53.89\% & 49.18\% & 42.97\% & 47.26\%          &                   & -                         & -                & -                         & -                         & -                & -                         \\ \cline{1-8} \cline{10-15} 
        & CC KNN           & 57.61\% & 48.22\% & 55.39\% & 54.82\% & 39.33\% & 47.90\%          &                   & -                         & -                & -                         & -                         & -                & -                         \\ \cline{1-8} \cline{10-15} 
        & BR AdaBoost DT                & 59.75\% & 45.73\% & 56.15\% & 52.96\% & 38.41\% & 48.22\%          &                   & -                         & -                & -                         & -                         & -                & -                    \\ \cline{1-8} \cline{10-15} 
        & BR Gradient T Boosting                 & 64.54\% & 45.72\% & 59.53\% & 55.37\% & 35.96\% & 47.76\%          &                   & -                         & -                & -                         & -                         & -                & -                  \\ \cline{1-8} \cline{10-15} 
        & BR Logistic Regression        & 58.71\% & 48.01\% & 56.13\% & 52.07\% & 42.03\% & 49.31\%          &                   & -                         & -                & -                         & -                         & -                & -                   \\ \cline{1-8} \cline{10-15} 
        & BR KNN       & 57.70\% & 48.23\% & 55.45\% & 54.14\% & 39.26\% & 47.62\% & & -                         & -                & -                         & -                         & -                & - \\ \cline{1-8} \cline{10-15} 
        & Adapted KNN           & 57.58\% & 48.08\% & 55.33\% & 53.64\% & 38.95\% & 47.25\% & & -                         & -                & -                         & -                         & -                & -                    \\ \cline{1-8} \cline{10-15}
\end{tabular}%
}
\end{table}

\begin{table}[htb]
\caption{Classification results for techniques prediction. Abbreviations: CC = Classifier Chain, BR = Binary relevance, DT = Decision Tree, T = Tree, NB = Naive Bayes}
\label{tabtechres}
\centering
\resizebox{\textwidth}{!}{%
\begin{tabular}{|ll|c|c|c|c|c|c|l|c|c|c|c|c|c|}
\cline{1-8} \cline{10-15}
\multicolumn{2}{|l|}{\multirow{3}{*}{}} & \multicolumn{6}{c|}{Without resampling}                                                                                                                                                      & \multirow{3}{*}{} & \multicolumn{6}{c|}{With resampling}                                                                                                                                                                  \\ \cline{3-8} \cline{10-15} 
\multicolumn{2}{|l|}{}                  & \multicolumn{3}{c|}{Micro}                                                                         & \multicolumn{3}{c|}{Macro}                                                              &                   & \multicolumn{3}{c|}{Micro}                                                                                  & \multicolumn{3}{c|}{Macro}                                                              \\ \cline{3-8} \cline{10-15} 
\multicolumn{2}{|l|}{}                  & Precision                             & Recall                      & $F_{0.5}$                         & Precision                   & Recall                      & $F_{0.5}$                        &                   & Precision                                      & Recall                      & $F_{0.5}$                         & Precision                   & Recall                      & $F_{0.5}$                        \\ \cline{1-8} \cline{10-15}
\multicolumn{2}{|l|}{Majority Baseline} & 9.57\%                                & 2.51\%                      & 6.11\%                       & 0.05\%                      & 0.48\%                      & 0.06\%                      &                   & -                                              & -                           & -                            & -                           & -                           & -                           \\ \hline
\multicolumn{15}{|l|}{Term Frequency}                                                                                                                                                                                                                                                                                                                                                                                                                              \\ \hline
        & CC Adaboost DT                & 36.64\%                               & 13.27\%                     & 27.05\%                      & 18.38\%                     & 8.98\%                      & 13.72\%                     &                   & -                                              & -                           & -                            & -                           & -                           & -                           \\ \cline{1-8} \cline{10-15} 
        & CC Bagging DT                 & 39.95\%                               & 5.83\%                      & 18.24\%                      & 18.50\%                     & 7.90\%                      & 12.32\%                     &                   & -                                              & -                           & -                            & -                           & -                           & -                           \\ \cline{1-8} \cline{10-15} 
        & CC Ridge Classifier           & 14.82\%                               & 25.59\%                     & 16.16\%                      & 10.88\%                     & 23.40\%                     & 11.66\%                     &                   & -                                              & -                           & -                            & -                           & -                           & -                           \\ \cline{1-8} \cline{10-15} 
        & CC Decision Tree              & 22.44\%                               & 17.94\%                     & 21.31\%                      & 18.64\%                     & 14.98\%                     & 16.46\%                     &                   & -                                              & -                           & -                            & -                           & -                           & -                           \\ \cline{1-8} \cline{10-15} 
        & BR AdaBoost DT                & 35.60\%                               & 16.04\%                     & 28.56\%                      & 18.23\%                     & 9.74\%                      & 14.23\%                     &                   & 26.44\%                                        & 23.02\%                     & 22.79\%                      & 14.65\%                     & 16.16\%                     & 12.45\%                     \\ \cline{1-8} \cline{10-15} 
        & BR Bagging DT                 & 39.30\%                               & 7.63\%                      & 21.42\%                      & 16.92\%                     & 7.53\%                      & 11.88\%                     &                   & 26.99\%                                        & 12.79\%                     & 16.81\%                      & 12.34\%                     & 15.43\%                     & 10.75\%                     \\ \cline{1-8} \cline{10-15} 
        & BR Gradient T Boosting        & 27.60\%                               & 12.67\%                     & 22.31\%                      & 20.25\%                     & 12.72\%                     & 16.26\%                     &                   & 18.47\%                                        & 20.32\%                     & 16.73\%                      & 14.80\%                     & 24.51\%                     & 13.96\%                     \\ \cline{1-8} \cline{10-15} 
        & BR Ridge Classifier           & 14.96\%                               & 25.96\%                     & 16.33\%                      & 10.98\%                     & 23.50\%                     & 11.74\%                     &                   & 11.40\%                                        & \textbf{31.83\%}            & 12.83\%                      & 9.05\%                      & \textbf{32.50\%}            & 10.04\%                     \\ \cline{1-8} \cline{10-15} 
        & BR Linear SVC                 & 22.87\%                               & 19.22\%                     & 21.98\%                      & 15.36\%                     & 11.39\%                     & 13.41\%                     &                   & 17.85\%                                        & 24.59\%                     & 17.97\%                      & 12.85\%                     & 17.01\%                     & 11.95\%                     \\ \cline{1-8} \cline{10-15} 
        & BR Decision Tree              & 23.06\%                               & 18.53\%                     & 21.94\%                      & 18.00\%                     & 14.76\%                     & 16.18\%                     &                   & 17.46\%                                        & 24.90\%                     & 17.11\%                      & 14.70\%                     & 21.66\%                     & 13.61\%                     \\ \cline{1-8} \cline{10-15}
        & Adapted Decision Tree                & 16.23\% & 14.06\% & 15.71\% & 9.99\% & 8.27\% & 9.00\%                     &                   & -                                              & -                           & -                            & -                           & -                           & -                           \\ \cline{1-8} \cline{10-15}
        & Adapted Extra Tree               & 12.86\% & 10.10\% & 12.19\% & 6.33\% & 4.95\% & 5.55\%                     &                   & -                                              & -                           & -                            & -                           & -                           & -                           \\ \hline
\multicolumn{15}{|l|}{Term Frequency-Inverse Document Frequency}                                                                                                                                                                                                                                                                                                                                                                                                   \\ \hline
        & CC Adaboost DT                & 37.06\%                               & 13.06\%                     & 26.98\%                      & 17.70\%                     & 8.44\%                      & 13.19\%                     &                   & -                                              & -                           & -                            & -                           & -                           & -                           \\ \cline{1-8} \cline{10-15} 
        & CC Decision Tree              & 23.18\%                               & 18.47\%                     & 22.02\%                      & 18.83\%                     & 14.85\%                     & 16.77\%                     &                   & -                                              & -                           & -                            & -                           & -                           & -                           \\ \cline{1-8} \cline{10-15} 
        & BR AdaBoost DT                & 35.04\%                               & 14.77\%                     & 27.41\%                      & 17.23\%                     & 9.05\%                      & 13.36\%                     &                   & 27.53\%                                        & 18.12\%                     & 22.32\%                      & 14.55\%                     & 13.00\%                     & 11.85\%                     \\ \cline{1-8} \cline{10-15} 
        & BR Gradient T Boosting        & 25.85\%                               & 11.60\%                     & 20.72\%                      & 18.83\%                     & 12.09\%                     & 15.21\%                     &                   & 19.45\%                                        & 16.76\%                     & 17.10\%                      & 15.33\%                     & 21.69\%                     & 14.38\%                     \\ \cline{1-8} \cline{10-15} 
        & Logistic Regression           & \multicolumn{1}{c|}{\textbf{52.82\%}} & \multicolumn{1}{c|}{3.66\%} & \multicolumn{1}{c|}{14.33\%} & \multicolumn{1}{c|}{7.86\%} & \multicolumn{1}{c|}{2.76\%} & \multicolumn{1}{c|}{5.18\%} &                   & \multicolumn{1}{c|}{\textit{\textbf{42.05\%}}} & \multicolumn{1}{c|}{4.98\%} & \multicolumn{1}{c|}{12.41\%} & \multicolumn{1}{c|}{7.53\%} & \multicolumn{1}{c|}{4.79\%} & \multicolumn{1}{c|}{5.64\%} \\ \cline{1-8} \cline{10-15} 
        & BR Perceptron                 & 30.45\%                               & 18.26\%                     & 26.82\%                      & 21.89\%                     & 14.61\%                     & 18.32\%                     &                   & 25.16\%                                        & 22.56\%                     & 23.03\%                      & 19.50\%                     & 21.29\%                     & 17.32\%                     \\ \cline{1-8} \cline{10-15} 
        & BR Linear SVC                 & 37.18\%                               & \textit{\textbf{29.79\%}}   & \textbf{35.02\%}             & \textbf{28.84\%}            & \textit{\textbf{22.67\%}}   & \textbf{25.06\%}            &                   & 31.16\%                                        & 32.86\%                     & \textit{\textbf{29.37\%}}    & \textit{\textbf{26.27\%}}   & 28.05\%                     & \textit{\textbf{22.87\%}}   \\ \cline{1-8} \cline{10-15} 
        & BR Decision Tree              & 20.72\%                               & 18.31\%                     & 20.15\%                      & 16.88\%                     & 14.57\%                     & 15.55\%                     &                   & 17.44\%                                        & 22.91\%                     & 17.02\%                      & 15.45\%                     & 20.21\%                     & 14.21\%                     \\ \cline{1-8} \cline{10-15}
        & Adapted Decision Tree                & 14.11\% & 12.19\% & 13.65\% & 8.14\% & 7.19\% & 7.44\%                     &                   & -                                              & -                           & -                            & -                           & -                           & -                           \\ \cline{1-8} \cline{10-15}
        & Adapted Extra Tree               & 12.54\% & 9.83\% & 11.85\% & 5.70\% & 4.68\% & 5.15\%                     &                   & -                                              & -                           & -                            & -                           & -                           & -                           \\ \hline
        \multicolumn{15}{|l|}{Word2Vec average}                                                                                                                                                                                                                                                                                      \\ \hline
        & CC Adaboost DT                & 29.88\% & 9.29\% & 20.67\% & 9.35\% & 4.05\% & 6.70\%          &                   & -                         & -                & -                         & -                         & -                & -                         \\ \cline{1-8} \cline{10-15} 
        & CC Linear SVC        & 36.21\% & 3.33\% & 11.70\% & 7.58\% & 3.60\% & 5.35\%          &                   & -                         & -                & -                         & -                         & -                & -                         \\ \cline{1-8} \cline{10-15} 
        & BR AdaBoost DT                & 26.88\% & 11.79\% & 21.36\% & 9.15\% & 4.82\% & 7.31\%          &                   & -                         & -                & -                         & -                         & -                & -                    \\ \cline{1-8} \cline{10-15} 
        & BR Perceptron       & 17.43\% & 15.88\% & 17.03\% & 8.33\% & 7.80\% & 6.37\% & & -                         & -                & -                         & -                         & -                & - \\ \cline{1-8} \cline{10-15} 
        & BR Bernoulli NB                 & 10.54\% & 47.77\% & 12.47\% & 5.21\% & 22.33\% & 6.06\% & & -                         & -                & -                         & -                         & -                & -                   \\ \cline{1-8} \cline{10-15}
        & BR Linear SVC                 & 49.01\% & 6.33\% & 20.70\% & 8.57\% & 3.27\% & 5.64\% & & -                         & -                & -                         & -                         & -                & -                   \\ \cline{1-8} \cline{10-15}
        & Adapted Extra Tree           & 12.53\% & 14.41\% & 12.85\% & 5.17\% & 5.75\% & 5.05\% & & -                         & -                & -                         & -                         & -                & -                    \\ \cline{1-8} \cline{10-15}
         \multicolumn{15}{|l|}{Word2Vec sum}                                                                                                                                                                                                                                                                                      \\ \hline
        & CC Adaboost DT                & 27.88\% & 7.34\% & 17.68\% & 8.77\% & 3.47\% & 6.14\%          &                   & -                         & -                & -                         & -                         & -                & -                         \\ \cline{1-8} \cline{10-15} 
        & BR AdaBoost DT                & 28.34\% & 11.69\% & 21.93\% & 9.63\% & 4.50\% & 7.30\%          &                   & -                         & -                & -                         & -                         & -                & -                    \\ \cline{1-8} \cline{10-15} 
        & BR Gradient T Boosting                 & 21.14\% & 8.41\% & 16.14\% & 7.38\% & 2.96\% & 5.00\%          &                   & -                         & -                & -                         & -                         & -                & -                  \\ \cline{1-8} \cline{10-15} 
        & BR Logistic Regression        & 20.49\% & 18.60\% & 19.98\% & 11.48\% & 10.53\% & 10.62\%          &                   & -                         & -                & -                         & -                         & -                & -                   \\ \cline{1-8} \cline{10-15}
        & BR Perceptron        & 16.39\% & 18.52\% & 16.77\% & 6.27\% & 7.96\% & 5.65\%          &                   & -                         & -                & -                         & -                         & -                & -                   \\ \cline{1-8} \cline{10-15}
        & BR Bernoulli NB        & 9.98\% & 49.20\% & 11.86\% & 5.10\% & 23.51\% & 5.93\%          &                   & -                         & -                & -                         & -                         & -                & -                   \\ \cline{1-8} \cline{10-15}
        & BR Linear SVC        & 11.64\% & 22.82\% & 12.87\% & 8.30\% & 15.49\% & 8.42\% & & -                         & -                & -                         & -                         & -                & - \\ \cline{1-8} \cline{10-15} 
        & Adapted Decision Tree           & 513.39\% & 14.23\% & 13.51\% & 5.63\% & 5.73\% & 5.42\% & & -                         & -                & -                         & -                         & -                & -                    \\ \cline{1-8} \cline{10-15}
\end{tabular}%
}
\end{table}

When we compare the text representation methods, we can observe that, overall, models using Word2Vec either with a sum or an averaging of the word vectors, underperformed compared to the ones using a TF or TF-IDF weighting system. We observed minor differences in the use of the average against the sum of the vector in the Word2Vec approach, but we cannot claim that one way is better than the other as the results depend on the employed classifier.

Considering the results of the different classifiers, we can see that adapted algorithms tend to underperform compared to the classifier chains and the binary relevance models, for both tactics and techniques predictions. Overall, the classification made with binary relevance instead of classifier chain performs slightly better, which means that the relationship between labels did not have as much impact as expected.

Some classifiers stand out among the others for each type of labels. For the classification by tactics, the AdaBoost Decision Tree, the Gradient Tree Boosting, Perceptron and the Linear SVC in classifier chains or binary relevance models have the best performance, independently from using TF-IDF or TF. The Bagging Decision tree, the Ridge classifier and the Logistic Regression perform well when used in binary relevance. They also perform well in a classifier chain if Logistic Regression and the Bagging Decision Tree classify text use TF weighting and the Ridge classifier use TF-IDF. Similar models work as well for techniques prediction, in addition to the Decision Tree classifier with binary relevance or classifier chain models.

The resampling on the tactics prediction resulted in an overall increase of the recall, but also a decrease in precision and, thus, in $F_{0.5}$ score. Only a few models (e.g. Extra Trees and Random Forest) improve because of the resampling. The Logistic Regression model does also increase its $F_{0.5}$ macro-averaged score when paired with TF-IDF, which is the maximum obtained in the resampling evaluation. On the techniques prediction, similarly, the recall increases, while the precision and the $F_{0.5}$ score decrease. The only improvements in performance are for models which performed very poorly without resampling.

Even though all metrics are essential, we want to give equal importance to each label. Thus looking at the macro-averaged $F_{0.5}$ scores, the binary relevance Linear SVC with a TF-IDF weighted text representation stands out compared to the other models, for both tactics and techniques predictions.

\section{Post-processing of classification results}\label{postprocessing}

From the analysis presented in the previous section, we select the binary relevance Linear SVC model with TF-IDF weighted bag-of-words.
To improve its results, we now try to take advantage of different properties of the ``Enterprise'' ATT\&CK framework such as the 
relationships between tactics and techniques. It is worth noting that, an effect of this property might have already been tested in classification models involving classifier chains. However, this issue does not concern the selected SVC model that can, therefore, benefit from this additional information.

\subsection{Undertaken approaches}

\subsubsection{Relationships between tactics and techniques}
Given the 
relationship between each technique and the tactic it belongs to, we use this property of the framework in the following ways:

\paragraph{Direct mapping from techniques to tactics}
This approach defines the tactics of a report based on the classification over the techniques. Because each technique belongs to one or more tactics, we can consider simple post-processing rules adding tactics to labels. For example, if a technique $Te_{1}$ is a label of a given report and $Te_{1}$ belongs to tactic $Ta_{1}$, then $Ta_{1}$ is also a label of the report.

\paragraph{Tactics as features}
Since the prediction of techniques underperformed the one on tactics, we can use the results from the classification by tactics to improve the one by techniques. In this case, we use the results of the tactics prediction as features for the classification by techniques. This is close to the classifier chain method, but it solely uses the tactics prediction to influence the techniques one. 

\paragraph{Confidence propagation}
Here, we want to use the method described in Ayoade \textit{et al.}~\cite{ayoade2018automated} (as well as Wu \textit{et al.}'s work~\cite{Wu2004Onotology}). The idea behind this approach is to use the confidence score of each technique and the confidence score of the associated tactics to create boosting factors. Each boosting factors are multiplied to each associated tactic confidence score, and this ensemble is added to the pre-existing technique confidence score. Based on the associated tactics confidence score, the techniques' confidence score increases or decreases, thus impacting the final predictions. To apply this method, it is worth noting that, Scikit-learn's Linear SVC required us to use the decision function scores instead of the confidence scores of the classifier as directly normalising the confidence scores (not automatically performed in the library) would have worsened model's performance.

\paragraph{Hanging node}

\begin{algorithm}[!htb]
\DontPrintSemicolon
\KwIn{\\$R$ \tcc{report}
$\left\{Ta_{i}^{R},...Te_{k}^{R}\right\}$ \tcc{all tactics and techniques predicted as being present in the $R$}
$Te_{x}$ \tcc{one of the techniques}
$Ta_{y}$ \tcc{one of the tactics}
$p(Te_{x} \in R)$ \tcc{the probability of $Te_{x}$ being predicted for $R$}
$p(Ta_{y} \in R)$ \tcc{the probability of $Ta_{y}$ being predicted for $R$}}
\KwOut{\\$\left\{Ta_{i}^{R},...Te_{k}^{R}\right\}$ \tcc{updated ensemble of tactics and techniques being present in the $R$}}
\KwData{\\$th\in \mathbb{R}$ \tcc{classification threshold}
$a, b, c, d \in \mathbb{R}$ \tcc{defined thresholds}}
\Begin{
\SetAlgoVlined
\If{$Te_{x} \rightarrow Ta_{y}$}{
\If{$p(Te_{x} \in R)>a>th$ and $b<p(Ta_{y} \in R)<th$}{
$\left\{Ta_{i}^{R},...Te_{k}^{R}\right\} += Ta_{y}$ \tcc{adding $Ta_{y}$ to the ensemble of tactics and techniques being present in the $R$}
}
\If{$th<p(Te_{x} \in R)<c$ and $p(Ta_{y} \in R)<d<th$}{
$\left\{Ta_{i}^{R},...Te_{k}^{R}\right\} -= Te_{x}$ \tcc{removing $Te_{x}$ to the ensemble of tactics and techniques being present in the $R$}
}
}
}
\caption{Hanging node approach\label{HN}}
\end{algorithm}

This approach is based on the observation that for 30\% of the techniques predicted in a report, not all related tactics were predicted, meaning that either the techniques or the tactics were incorrectly identified. The analysis of the distribution of the confidences' frequency shows that false predictions tend to be closer to the ``successful'' classification threshold (especially when it comes to tactics). The Hanging node approach tries to leverage this aspect by considering all confidence scores when adding tactics or removing techniques from the labels. As presented in Algorithm~\ref{HN}, for a connected pair ``tactic-technique'', if the technique is predicted with a high confidence score, while the tactic is not predicted, but with a confidence score close to the classification threshold $th$, then we can add the tactic to the tactics and techniques labeling the report. On the contrary, if the techniques are present, with a confidence score near to the classification threshold $th$ and the tactic is not predicted for the report, with a low confidence score, then we can remove the techniques from the predicted labels. The thresholds $a, b, c, d$ were defined after testing different values and comparing the improvement in classification performance.

According to our tests, a classification threshold of $th~=~0.5$ and thresholds $a~=~0.55, b~=~0.05, c~=~0.95, d~=~0.30$, allow the highest macro-averaged $F_{0.5}$ score. As we use the Linear SVC from Scikit-learn, we defined the confidence score by scaling the decision function scores using min-max scaling with $min~=~-1$ and $max~=~1$.

Further similar strategies based on relationships between techniques and tactics are difficult to implement, as several techniques correspond to the same tactics. Adding a technique if a related tactic is also present in the report would be misguided, as the high probability could be due to another technique. Similarly, removing a tactic, if a related technique is absent, would be disadvantageous for basically the same reason.

\subsubsection{Relationships between techniques}
Always based on the ATT\&CK framework structure, we can compute joint probabilities between a couple of techniques based on their common appearances within the same malware, tool or group. Using these probabilities, we decided to test three different methods to improve techniques predictions.

\paragraph{Rare association rules}
This approach follows the work of Benites and Sapozhnikova~\cite{benites2015improving}, based on a selection of association rules between techniques. The first step is to calculate the Kulczynski measure~\cite{kulczinski1927die} for each pair of techniques. These values are forming a curve from which we can determine the variance. If this variance is low, the threshold to decide on the pairing rules based on their Kulczynski measure is based on the median of differences between neighbour values. If this variance is high, it is based on the average of the values slightly lower than the mean of this curve.
\paragraph{Steiner tree association rules}
This approach has been described by Soni \textit{et al.}~\cite{soni2017post}, and focuses on a formulation of the label coherence as a Steiner Tree Approximation problem~\cite{gilbert1968steiner}. Once the techniques prediction is performed, for each report, we create a directed tree~\cite{mehlhorn1988faster} in which edges have weights corresponding to the conditional probabilities between the two nodes and the direction is given by the following criterion:

{\centering \[\left\{\begin{matrix}
Te_{i}\rightarrow Te_{j}, & p(Te_{i}|Te_{j}) \leq p(Te_{j}|Te_{i}),\\ 
Te_{i}\leftarrow Te_{j}, & otherwise.
\end{matrix}\right.\]\par }

Then, we use Edmond's algorithm~\cite{edmonds1967optimum} to obtain a reduced tree and a limited number of connections between techniques. Based on the predictions of the classification, we search the graph for techniques which descend from the predicted techniques with the $K$ highest weights. In our tests, we found that $K~=~15$ is the value maximizing the success of this pre-processing strategy.

\paragraph{Knapsack}
This approach also comes from the paper of Soni \textit{et al.}~\cite{soni2017post} but, this time, the authors consider the label assignment as a resource allocation problem. In this case, we solve the 0-1 Knapsack problem~\cite{martello1990knapsack}, in which a new label is selected if its conditional probability (based on the predicted labels) increases the overall log-likelihood.

\subsection{Results and discussion}

Table~\ref{tab3} shows the final results of all our post-processing approaches. To better investigate the effects of post-processing approaches we also applied those assuming a perfect match of techniques (when predicting the tactics) and vice-versa (Table~\ref{tab3perf}). For all results, we use a new baseline given by the pure classification without any post-processing.

\begin{table}[htb]
\caption{Comparison of post-processing approaches for tactics and techniques.}
\label{tab3}
\centering
\resizebox{\textwidth}{!}{%
\begin{tabular}{|ll|l|l|l|l|l|l|}
\hline
\multicolumn{2}{|l|}{\multirow{2}{*}{Approaches}} & \multicolumn{3}{c|}{Micro}                                                                                                       & \multicolumn{3}{c|}{Macro}                                                                                                        \\ \cline{3-8} 
\multicolumn{2}{|l|}{}                         & \multicolumn{1}{c|}{Precision}                                    & \multicolumn{1}{c|}{Recall}         & \multicolumn{1}{c|}{$F_{0.5}$}                   & \multicolumn{1}{c|}{Precision}               & \multicolumn{1}{c|}{Recall}         & \multicolumn{1}{c|}{$F_{0.5}$}                    \\ \hline
\multicolumn{8}{|l|}{Tactics}                                                                                                                                                                                                                                                                                         \\ \hline
                    & Inde.                    & \multicolumn{1}{c|}{\textbf{65.64\% \tiny$\pm$3.76\%}} & \multicolumn{1}{c|}{64.69\% \tiny$\pm$3\%} & \multicolumn{1}{c|}{\textbf{65.38\% \tiny$\pm$2.87\%}} & \multicolumn{1}{c|}{\textbf{60.26\% \tiny$\pm$3.2\%}} & \multicolumn{1}{c|}{58.50\% \tiny$\pm$3.68\%} & \multicolumn{1}{c|}{\textbf{59.47\% \tiny$\pm$2.29\%}} \\ \hline
                    & 1.a.                     & 63.32\% \tiny$\pm$6.42\%                               & 52.34\% \tiny$\pm$5.08\%                      & 60.45\% \tiny$\pm$3.8\%                              & 57.49\% \tiny$\pm$5.41\%                               & 47.94\% \tiny$\pm$5.87\%                      & 54.59\% \tiny$\pm$3\%                               \\ \hline
                    & 1.d.                     & 59.60\% \tiny$\pm$2.89\%                                & \textbf{68.04\% \tiny$\pm$3.06\% }             & 61.08\% \tiny$\pm$3.42\%                               & 54.41\% \tiny$\pm$2.97\%                                & \textbf{61.28\% \tiny$\pm$3.12\% }             & 55.42\% \tiny$\pm$3.2\%                                \\ \hline
\multicolumn{8}{|l|}{Techniques}                                                                                                                                                                                                                                                                                      \\ \hline
                    & Inde.                    & \textbf{37.18\% \tiny$\pm$6.75\%}                               & 29.79\% \tiny$\pm$5.91\%                      & \textbf{35.02\% \tiny$\pm$5.32\%}                              & 28.84\% \tiny$\pm$6.9\%                               & 22.67\% \tiny$\pm$5.94\%                      & 25.06\% \tiny$\pm$6.09\%                               \\ \hline
                    & 1.b.                     & 31.55\% \tiny$\pm$5\%                               & 35.17\% \tiny$\pm$6\%                      & 31.97\% \tiny$\pm$4.31\%                              & 24.70\% \tiny$\pm$6.29\%                               & 22.74\% \tiny$\pm$6.3\%                      & 22.38\% \tiny$\pm$5.72\%                               \\ \hline
                    & 1.c.                     & 33.07\% \tiny$\pm$7.11\%                               & \textbf{38.17\% \tiny$\pm$5.41\%}             & 33.69\% \tiny$\pm$6.2\%                              & 28.14\% \tiny$\pm$6.72\%                               & \textbf{28.19\% \tiny$\pm$4.99\%}             & 26.06\% \tiny$\pm$5.69\%                               \\ \hline
                    & 1.d.                     & 32.19\% \tiny$\pm$6.05\%                               & 29.27\% \tiny$\pm$6.2\%                      & 31.34\% \tiny$\pm$5.23\%                               & \textbf{32.35\% \tiny$\pm$6.68\% }                      & 22.21\% \tiny$\pm$4.89\%                       & \textbf{27.52\% \tiny$\pm$6.03\% }                      \\ \hline
                    & 2.a.                     & 33.70\% \tiny$\pm$6.79\%                      & 36.26\% \tiny$\pm$5.16\%                      & 33.89\% \tiny$\pm$5.76\%                              & 28.08\% \tiny$\pm$6.8\%                               & 25.18\% \tiny$\pm$5.22\%                      & 25.20\% \tiny$\pm$5.78\%                               \\ \hline
                    & 2.b.                     & 37.06\% \tiny$\pm$6.77\%                               & 29.79\% \tiny$\pm$5.99\%                      & 34.93\% \tiny$\pm$5.34\%                     & 28.84\% \tiny$\pm$6.94\%                               & 22.67\% \tiny$\pm$5.91\%                      & 25.06\% \tiny$\pm$6.11\%                               \\ \hline
                    & 2.c.                     & 33.98\% \tiny$\pm$5.92\%                               & 33.88\% \tiny$\pm$6\%                      & 33.68\% \tiny$\pm$\%5.02                              & 28.38\% \tiny$\pm$6.88\%                               & 23.60\% \tiny$\pm$5.9\%                      & 24.80\% \tiny$\pm$6.06\%                               \\ \hline
\end{tabular}%
}
\end{table}

\begin{table}[htb]
\centering
\caption{Comparison of post-processing approaches for tactics and techniques (considering a perfect techniques and tactics predictions respectively).}
\label{tab3perf}
\resizebox{\textwidth}{!}{%
\begin{tabular}{|ll|l|l|l|l|l|l|}
\hline
\multicolumn{2}{|l|}{\multirow{2}{*}{Approaches}} & \multicolumn{3}{c|}{Micro}                                                                                     & \multicolumn{3}{c|}{Macro}                                                                                      \\ \cline{3-8} 
\multicolumn{2}{|l|}{}                         & \multicolumn{1}{c|}{Precision}      & \multicolumn{1}{c|}{Recall}         & \multicolumn{1}{c|}{$F_{0.5}$}          & \multicolumn{1}{c|}{Precision}      & \multicolumn{1}{c|}{Recall}         & \multicolumn{1}{c|}{$F_{0.5}$}           \\ \hline
\multicolumn{8}{|l|}{Tactics}                                                                                                                                                                                                                                                     \\ \hline
                    & Inde.                    & \multicolumn{1}{c|}{65.64\% \tiny$\pm$3.76\%} & \multicolumn{1}{c|}{64.69\% \tiny$\pm$3\%} & \multicolumn{1}{c|}{65.38\% \tiny$\pm$2.87\%} & \multicolumn{1}{c|}{60.26\% \tiny$\pm$3.2\%} & \multicolumn{1}{c|}{58.50\% \tiny$\pm$3.68\%} & \multicolumn{1}{c|}{59.47\% \tiny$\pm$2.29\%} \\ \hline
                    & 1.a.                     & \textbf{100\% \tiny$\pm$0\% }                & \textbf{100\% \tiny$\pm$0\% }                & \textbf{100\% \tiny$\pm$0\% }               & \textbf{100\% \tiny$\pm$0\% }                & \textbf{100\% \tiny$\pm$0\% }                & \textbf{100\% \tiny$\pm$0\% }                \\ \hline
                    & 1.d.                     & 68.09\% \tiny$\pm$4.02\%                      & 72.37\% \tiny$\pm$2.97\%                      & 68.84\% \tiny$\pm$3.24\%                     & 63.32\% \tiny$\pm$3.1\%                      & 66.29\% \tiny$\pm$4\%                      & 63.54\% \tiny$\pm$2.08\%                      \\ \hline
\multicolumn{8}{|l|}{Techniques}                                                                                                                                                                                                                                                  \\ \hline
                    & Inde.                    & 37.18\% \tiny$\pm$6.75\%                               & 29.79\% \tiny$\pm$5.91\%                      & 35.02\% \tiny$\pm$5.32\%                              & 28.84\% \tiny$\pm$6.9\%                               & 22.67\% \tiny$\pm$5.94\%                      & 25.06\% \tiny$\pm$6.09\%                               \\ \hline
                    & 1.b.                     & \textbf{51.54\% \tiny$\pm$5.01\%}             & \textbf{55.90\% \tiny$\pm$4.23\%}             & \textbf{52.24\% \tiny$\pm$4.4\%}            & \textbf{41.05\% \tiny$\pm$3.94\%}             & \textbf{39.04\% \tiny$\pm$5.13\%}             & \textbf{38.35\% \tiny$\pm$3.16\%}             \\ \hline
                    & 1.c.                     & 36.51\% \tiny$\pm$8\%                      & 48.75\% \tiny$\pm$4.13\%                      & 38.14\% \tiny$\pm$7.54\%                     & 34.21\% \tiny$\pm$7.43\%                      & 37.12\% \tiny$\pm$4.13\%                      & 32.36\% \tiny$\pm$6.24\%                      \\ \hline
                    & 1.d.                     & 38.40\% \tiny$\pm$7\%                      & 28.39\% \tiny$\pm$6.14\%                      & 35.44\% \tiny$\pm$5.42\%                     & 29.14\% \tiny$\pm$7.1\%                      & 22.40\% \tiny$\pm$6.22\%                      & 25.23\% \tiny$\pm$6.37\%                      \\ \hline
\end{tabular}%
}
\end{table}

When it comes to tactics prediction, the independent classification is the best. This might be related to the fact that techniques predictions had worse results than tactics ones and thus any post-processing depending on these last could decrease results on the tactics. However, considering a perfect techniques prediction, both post-processing approaches would help improve the tactics classification. The approach ``\textit{Direct mapping from techniques to tactics}" would have a perfect performance, as it is purely rule-based. However, it is unlikely that the techniques prediction would improve without the tactic prediction improving. Because of that, approach ``\textit{Hanging node}" seems to be the most promising approach to improve the independent classification.

For what concerns techniques prediction, the independent classification started with a lower baseline. The use of the approaches ``\textit{Confidence propagation}" and ``\textit{Hanging node}" are likely possible improvements. In all cases, except for the approach ``\textit{Hanging node}", the $F_{0.5}$ score changed mainly due to lower precision and increased recall related to the addition of techniques to the predicted set. In this case, the $F_{0.5}$ score increasing in ``\textit{Hanging node}" is especially interesting, as it is due to the decrease of false-positive and, thus, the increase of the precision.

Approaches relying on the relationship between techniques are close to the classification without post-processing. From these results, we conclude that these approaches might not well adapt to our problem or to the data we use. Approach ``\textit{Rare association rules}" could probably fit better in a hierarchical environment with known conditional probabilities. It is also possible that the ground truth we used is incomplete, as, even though it is based on data collected and analysed by experts, it represents only a sample of malware, tools and campaigns being observed in the past years.

If tactics prediction were perfect, and without any change in the prediction of techniques, approach ``\textit{Tactics as features}" would have the highest score for all metrics. Once again, we cannot choose an approach based on this test and the results from Table~\ref{tab3perf}, since the techniques prediction would probably improve with the perfect tactics prediction. However, since all approaches are better or equal than the independent classification, we conclude that post-processing is needed for techniques prediction. 
It is worth noting that, approach ``\textit{Hanging node}" is barely higher than the independent classification compared to ``\textit{Confidence propagation}", while the opposite is true in the actual tests.
Approach ``\textit{Hanging node}" also presents worse results when the tactics prediction is perfect. For this reason, we conclude that approach ``\textit{Confidence propagation}" might likely overperform ``\textit{Hanging node}".
Furthermore, ``\textit{Hanging node}" and ``\textit{Confidence propagation}" approaches carry on very different strategies and, from our testing, they seem targeting different types of techniques. ``\textit{Hanging node}" seems to affect very little, but efficiently, the results from the independent classification and only modifies results of the techniques with the highest number of reports. ``\textit{Confidence propagation}" targets any techniques and modify many predictions, which would potentially improve the classification for the hard-to-predict techniques. However, from our observation, this approach succeeds almost as often as it fails, thus worsening the prediction of some labels while improving the prediction of others. 

At the end of the evaluation, we are not in a position to draw conclusive proves for precisely ranking the described post-processing methods. Based on the analysis of the results, not using any post-processing method for the tactics classification is probably a safe choice. For what concerns the techniques, we believe that one among ``\textit{Confidence propagation}" and ``\textit{Hanging node}" would likely improve the independent classification. 
In what follows, within our tool, we decided to implement both approaches but to use ``\textit{Hanging node}" as default due to its better results in our current dataset. However, in case of retraining on a new dataset, the tool will be compare results coming from the two approaches and dynamically choose the one that best perform.

\section{rcATT}\label{tool}

\begin{figure}[!thb]
        \center{\includegraphics[width=9.5cm]
        {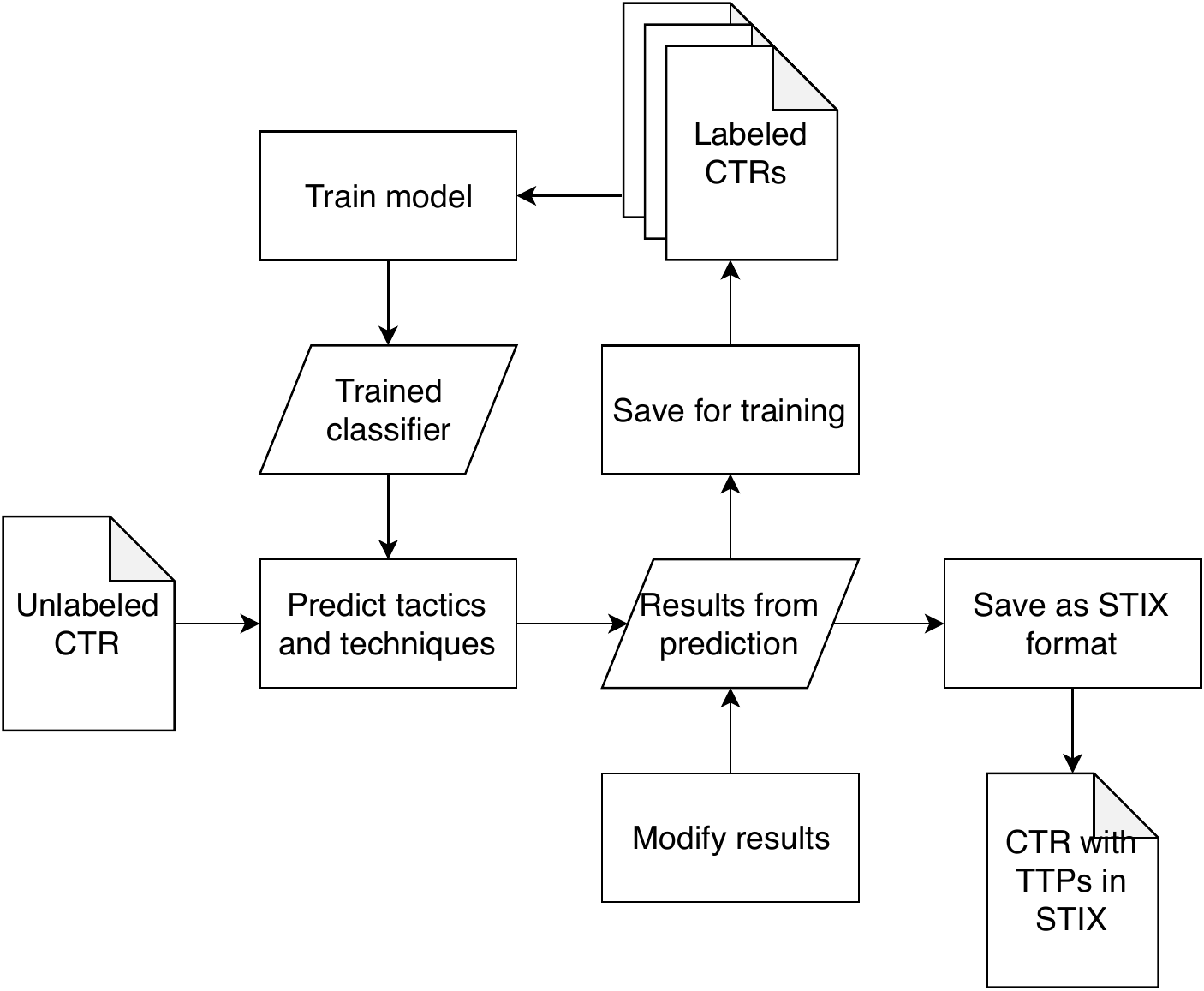}}
        \caption{\label{fig2} Organisation of rcATT.}
\end{figure}

One important contribution of our research was to implement our findings in a tool that would allow security experts to automate the analysis of CTRs. We call our tool rcATT, as for ``reports classification by adversarial tactics and techniques". Based on the previous two sections, we decided to implement a classification that uses a TF-IDF weighted bag-of-words text representation with a binary relevance Linear SVC. This classification is followed with a post-processing performed either via the hanging node approach or the confidence propagation one (depending on the training data).

Figure~\ref{fig2} illustrate tools' main functional building blocks. From a cyber threat report, in a text format, entered by the users, our trained model predicts different tactics and techniques. The user will be able to visualize the results and the confidence score associated with each prediction, either based on the Min-Max scaling of the decision function score from the classifier or the re-evaluation score, based on the post-processing method. When users do not agree with the results, they can change them. In all cases, new results can be saved with the old training data (the original labelled cyber threat reports) so to be used to train the classifier again. The retraining of the classifier must be activated manually by the user. This is done to avoid automated retraining that might slow down the tool or mistakenly involve unwanted data. As already mentioned, during the retraining, the tool also autonomously choose which post-processing method to use to maximize predictions results. This approach involves also the dynamic definition of specific thresholds (e.g., the one related to the hanging node methods). 

\begin{figure}[htb]
        \center{\includegraphics[width=\columnwidth]
        {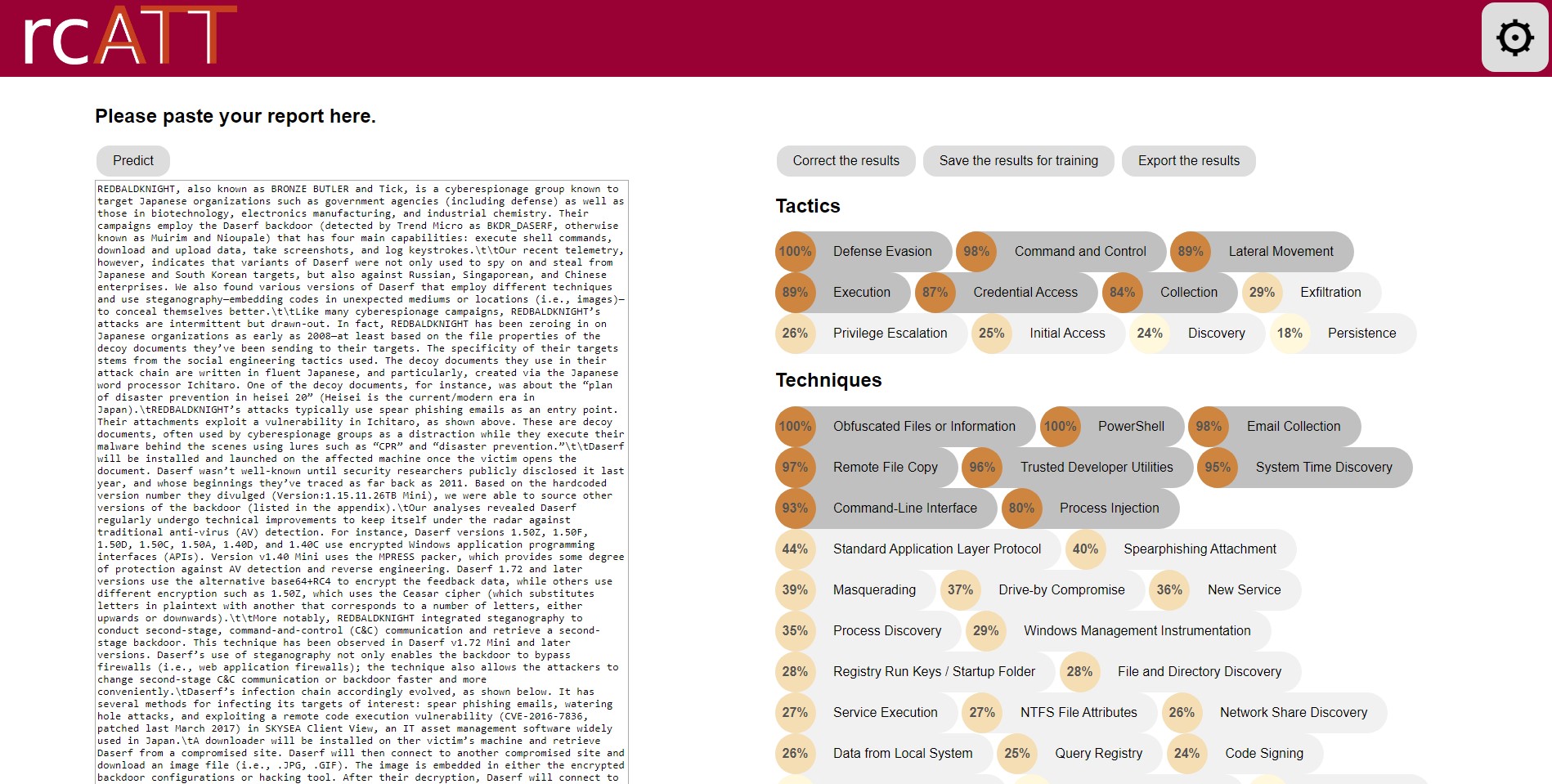}}
        \caption{\label{fig4} rcATT graphical interface.}
\end{figure}

Classification results can be saved in a JSON file using STIX format. This later allows any other tool to easily access all identified tactics and techniques. General information about the predictions is exported in a STIX ``Report" object (e.g., title of the original report, dates, etc.). The ``description" property of the report object represents the original content of the CTR. Most important, all prediction results (as well as any prediction added by the users) are available using the specific identification numbers available within the STIX version of the ATT\&CK framework and linked to the Report object using the ``object\_refs" property.

Finally, the tool works with two graphic modes: command-line and graphical interface (Figure \ref{fig4}). The command-line version of the tool is meant to be efficient to use (e.g., cases in which the user want to predict TTPs from a large number of CTRs). The graphical interface is interactive and allows users to modify classification results and save the changes to the training set and thus give feedback to the tool.

\section{Related work and comparison with similar approaches}\label{comparison}
The vast majority of existing scientific work on the automated analysis of (unstructured) CTRs concerns the extraction of IOCs~\cite{liao2016acing,zhu2018chainsmith}. Further related work looks at other unstructured sources, like hacker forums, to extract insights on adversarial tools~\cite{deliu2018collecting}, or on more general security vulnerabilities~\cite{mulwad2011extracting} and related concepts~\cite{jones2015towards,ramnani2017semi}.

Only few research papers aim to retrieve TTPs in unstructured data, as we do. One of them was written by Zhu \textit{et al.}~\cite{zhu2016featuresmith} and presents FeatureSmith, a tool that mines the security literature to learn features for training machine learning classifiers to detect Android malware. FeatureSmith is specifically tailored to the task of malware detection and cannot be easily generalized to other other settings. In contrast, our focus lies on the automated extraction of general ATT\&CK tactics and techniques.

TTPDrill~\cite{husari2017ttpdrill} and ActionMiner~\cite{husari2018using}, developed by Husari \textit{et al.}, but also the tool by Ayoade \textit{et al.}~\cite{ayoade2018automated} are the closest tools to ours, as they extract threat actions from CTRs in order to link them to specific ATT\&CK tactics and techniques. The biggest problem with those works is, however, that their achieved results cannot be reproduced by the scientific community at large as their used datasets or their tools are not openly available. In fact, \cite{ayoade2018automated} reports a big difference in accuracy when running TTPDrill~\cite{husari2017ttpdrill} on their own (closed) datasets in comparison to the accuracy achieved on the (closed) dataset as reported in~\cite{husari2017ttpdrill}. Due to the closed source nature of these works, these inconsistencies cannot be explained and motivate our open study on the extraction of TTPs from CTRs. In comparison to all related work, our approach is described for easy reproducibility by only using open data and by making our final tool publicly available as open source.

On the conceptual level, however, there are also some key differences between the above-stated approaches and ours. First of all, TTPDrill~\cite{husari2017ttpdrill} uses a different classification approach without a clear separation among tactics and techniques (treated as similar labels). In this regard, Ayoade \textit{et al.}~\cite{ayoade2018automated} describe a more similar approach, clearly separating tactics and techniques, despite not implementing a fully-contained tool in the end. Both tools employ different classifier and/or pre-processing approaches with respect to the ones implemented within rcATT. Finally, none of the two works allow users to feedback results back to the tool making any improvement impossible. Again, due to the lack of open-source code or used datasets (as well as any detailed description on how to re-implement their solutions), any further comparison on classification results is not possible.

A different tool, called Unfetter Insight, was instead available on Github~\cite{unfetter_insight}. Despite lacking a comprehensive description (e.g., publication) that allows a qualitative analysis with respect to rcATT, we could in this case carry out a quantitative comparisong between the classification results. Unfetter Insight has been developed to detect possible ATT\&CK tactics and techniques within reports in TXT, PDF or HTML format.  Based on our understanding of the code, the detection is implemented by the Babelfish software created by W. Kinsman in 2017. This software learns topics from a wiki-like corpus and retrieve the presence of those topics in a document. The tool creates text convolutions from unknown documents, identifying those having a good overlap with the ones presented in the corpus. Based on the term frequency of the vocabulary in the selected convolutions, tactics and techniques are predicted using a binary relevance Multinomial Naive Bayes classifier.
It is finally worth noting that Unfetter Insight assumptions for the models are quite different with respect to the ones used in rcATT. In fact, we base our classification on a training set of labelled reports while they use a dictionary-like learning set. For this reason, we could not add our training set without compromising the concept behind Babelfish. Therefore, we decided to proceed with a comparison by using their training set within both solutions\footnote{Interestingly, their wiki-like corpus is, in fact, the content of the ATT\&CK website with the techniques definitions, and thus a subset of our dataset.}.

Table~\ref{tab2} presents the classification results of rcATT and Unfetter Insight on identical training set and dataset. The table shows that Unfetter Insight underperforms in the majority of the proposed metrics and overperforms only in the one we considered the least important in our first assumptions.

\begin{table}[]
\caption{Comparison between rcATT and Unfetter Insight.}
\centering
\resizebox{9.5cm}{!}{%
\begin{tabular}{|ll|c|c|c|c|c|c|}
\hline
\multicolumn{2}{|l|}{\multirow{2}{*}{Tools}}                       & \multicolumn{3}{c|}{Micro}                             & \multicolumn{3}{c|}{Macro}                             \\ \cline{3-8} 
\multicolumn{2}{|l|}{}                                        & Precision        & Recall           & $F_{0.5}$             & Precision        & Recall           & $F_{0.5}$             \\ \hline
\multicolumn{8}{|l|}{Tactics}                                                                                                                                                   \\ \hline
 & rcATT                                                      & \textbf{79.31\%} & 12.20\%          & \textbf{37.75\%} & \textbf{81.73\%} & 8.21\%           & \textbf{23.23\%}  \\ \hline
 & \begin{tabular}[c]{@{}l@{}}Unfetter\\ insight\end{tabular} & 19.09\%          & \textbf{26.76\%} & 20.25\%          & 19.17\%          & \textbf{23.14\%} & 19.28\%          \\ \hline
\multicolumn{8}{|l|}{Techniques}                                                                                                                                                \\ \hline
 & rcATT                                                      & \textbf{72.22\%} & \textbf{2.07\%}  & \textbf{9.30\%}  & \textbf{20.60\%} & \textbf{4.33\%}  & \textbf{10.11\%} \\ \hline
 & \begin{tabular}[c]{@{}l@{}}Unfetter\\ insight\end{tabular} & 3.54\%           & 1.11\%           & 2.46\%           & 1.31\%           & 0.64\%           & 0.74\%           \\ \hline
\end{tabular}%
}
\label{tab2}
\end{table}
\section{Conclusion}

With its constant growth, CTI sharing has become an essential activity of the security lifecycle. Despite this, security teams around the world still face the challenge of handling huge amount of threat intelligence data. Above all, dealing with unstructured data remain a cumbersome and time-consuming task.  

The work presented in this paper aims at simplifying this task by enabling automated extraction of valuable CTI information, namely Tactics, Techniques and Procedures (TTPs), from textual cyber security reports. Our approach takes a cue from the ones of Ayoade \textit{et al.}~\cite{ayoade2018automated} and Husari \textit{et al.}~\cite{husari2017ttpdrill} and investigates text multi-label classification techniques that can fulfill the aforementioned goal. Additionally, we propose approaches for post-processing classification results and improving classifiers' overall performance in terms of precision and recall. By taking advantage of our training dataset, the MITRE ATT\&CK framework, we show that 
relationships among the chosen classification labels, the framework's tactics and techniques, can help refining the classification and decreasing the number false-positives. 

Based on our findings, we developed rcATT, an interactive tool for the automated analysis of cyber threat reports. The tool aims at supporting security experts to dig into human-readable data and extract ATT\&CK tactics and techniques that most likely describe the content of the text. rcATT gives users the possibility to dinamically adjust or fix the classification results and, consequently, feedback the internal classifier with correct information, improving its performance over time.

The tool has been tested and compared with state-of-the-art approaches discussed in literature or already available as open-source solutions. Results of these comparisons show rcATT competitiveness and validate the research directions foreseen within our work. Finally, rcATT has been recently integrated into tool-chains of real operational corporate Computer Emergency Response Teams (CERTs) to support Threat Intelligence sharing and analytics processes. This last experience further confirms the importance of  researches of this kind and their likely impact in broader contexts such as automated incident handling and response.

\bibliographystyle{splncs04}
\bibliography{TTPretrieval}

\end{document}